\documentclass[12pt]{article}
\usepackage{a4wide,amssymb,cite}
\pdfoutput=1

\usepackage{a4wide,amssymb,graphicx}
\usepackage{epsfig}
\usepackage[usenames,dvipsnames]{color}
\usepackage{slashed}
\usepackage{amssymb,cite,graphicx}
\usepackage{slashed}
\usepackage{amsmath,bm,bbm}
\usepackage{amsfonts}
\usepackage[titletoc,title]{appendix}
\usepackage[small]{caption}
\usepackage[margin=1in]{geometry}
\usepackage[multiple]{footmisc}
\usepackage{mathtools}
\usepackage{slashed}
\usepackage[nottoc]{tocbibind}
\usepackage{xcolor}

\newcommand{\be}{\begin{equation}}
\newcommand{\ee}{\end{equation}}
\newcommand{\bea}{\begin{eqnarray}}
\newcommand{\eea}{\end{eqnarray}}

\def\circa#1{\,\raise.3ex\hbox{$#1$\kern-.75em\lower1ex\hbox{$\sim$}}\,}

\begin{document}

\begin{titlepage}
%
%


%

\begin{centering}
\vspace{1cm}
{\Large {\bf Higgs-like pole inflation with loop corrections\vspace{0.2cm}\\ in light of ACT results}} \\

\vspace{1.5cm}

\begin{centering}
{\bf  Jeonghak Han$^\star$, Hyun Min Lee$^\dagger$, and Jun-Ho Song$^\sharp$}
\end{centering}
\\
\vspace{.5cm}

{\it Department of Physics, Chung-Ang University, Seoul 06974, Korea.}

\vspace{.5cm}


\end{centering}
\vspace{2cm}

\begin{abstract}
\noindent
We present the Coleman-Weinberg potential for the inflaton in the pole inflation scenarios such as the Higgs pole inflation and the Peccei-Quinn (PQ) pole inflation. The loop corrections stem from the Standard Model particles and extra singlet scalar fields in the former case, making the quartic coupling for the Higgs inflaton modified by the inflaton-dependent power corrections during inflation. We also obtain similar power corrections to the quartic coupling for the PQ inflaton, depending on the realizations of the PQ symmetry in KSVZ and DFSZ models.
We show that the loop corrections can shift the spectral index in the pole inflation to a larger value in favor of the ACT results, while being compatible with the bound on the tensor-to-scalar ratio. For a positive one-loop beta function for the inflaton quartic coupling (namely, $b_1>0$), a sub-dominant contribution from the two-loop corrections can be accommodated. On the other hand, if the one-loop beta function for the inflaton coupling is negative (namely, $b_1<0$), we need sizable contributions from two-loops that are larger than the one-loop corrections due to the ACT results.

\end{abstract}

\vspace{2.5cm}

\begin{flushleft} 
$^\star$Email: jeonghakhan@gmail.com \\
$^\dagger$Email: hminlee@cau.ac.kr \\
$^\sharp$Email: thdwnsgh1003@gmail.com
\end{flushleft}

\end{titlepage}

\section{Introduction}

Cosmic inflation is introduced to solve various problems in the Standard Big Bang cosmology such as horizon, homogeneity, isotropy, flatness, etc. The inflaton scalar field is required to derive a slow-roll inflation just after Big Bang, and it makes a smooth transition of the universe to the period of radiation-domination for Big Bang nucleosynthesis possible with a reheating process. The measured red-tilt of the power spectrum of the Cosmic Microwave Background (CMB) isotropies and the potential discovery of the primordial gravitational waves in the future CMB experiments could provide clues to reconstructing the shape of the inflaton potential\cite{review}. Moreover, the precise CMB measurements enable us to determine the cosmological parameters such as the fractions of dark matter and dark energy in the total energy density of the universe in combination of other large-scale measurements \cite{Planck}.

Recently, there has been an announcement of the precise measurements of CMB from Atacama Cosmology Telescope (ACT), in particular, at small scales, so a combined analysis of the Planck and ACT data has hinted at an increase in the spectral index as compared to Planck data \cite{ACT}, showing a tension between some of the well-studied inflationary predictions and the observations. Higgs inflation \cite{Higgsinflation} and Starobinsky inflation \cite{Starobinsky:1980te} have drawn a lot of attention, because they are based on simple non-minimal gravity interactions and the bound on the tensor-to-scalar ratio from Planck and Keck \cite{keck} is easily satisfied. However, in view of the ACT results, it is tempting to make modifications in the previous inflation models \cite{loops} or investigate the details of the reheating dynamics for a larger number of efoldings \cite{reheating} or suggest new inflation models \cite{models}.

In this article, we consider the Coleman-Weinberg (CW) potential and its effects on the inflationary predictions in a class of the pole inflation models where the inflaton is conformally coupled to gravity \cite{poleinflation,Lee:2021dgi,Aoki:2022dzd}. The pole inflation takes place near the pole of the inflaton kinetic term in the Einstein frame and it can be realized in particle physics models where the role of the inflation is played by the Standard Model (SM) Higgs field \cite{Clery:2023ptm} or the Peccei-Quinn (PQ) field in the extension of the SM with a $U(1)$ PQ symmetry \cite{Lee:2023dtw,Lee:2024bij}. It is remarkable that the Higgs pole inflation can be attributed to the Weyl gravity with an extended custodial symmetry in the scalar sector including the dilaton \cite{Lee:2024rjw}.

Considering the renormalization group (RG) improved inflaton potential in both Higgs and PQ inflation models, we parametrize the effects of the loop corrections as a running quartic coupling for the inflaton at the two-loop order. Then, we analyze the impacts of the loop corrections on the spectral index and the tensor-to-scalar ratio in the perturbative regime where the running quartic coupling does not change significantly during inflation. As a result, we discuss the implications of the ACT results for the beta function coefficients for the inflaton quartic coupling and how the couplings of the inflaton to new particles or the content of the models are constrained. There are previous studies on the RG improved inflaton potential in Higgs-like inflation models \cite{loops2}.

The paper is organized as follows.
We begin with the description of the setup for the pole inflation and show how the inflaton potential gets modified in the presence of the one-loop CW potential coming from the SM or the extra fields in both Higgs pole inflation and PQ pole inflation. Then, we parametrize the loop corrections as a running quartic coupling for the inflaton and discuss how the inflationary predictions vary depending on the beta function coefficients of the running quartic coupling at the two-loop level. Finally, conclusions are drawn.
There is one appendix dealing with the details of the derivation of the renormalization group equations.

\section{Setup}

We first describe the setup for the pole inflation with complex scalar fields at tree level, applicable to the rest of the sections devoted to the loop corrections.

We take the inflaton to be conformally coupled to gravity in the leading order, while the conformal symmetry can be broken by the Planck mass, the mass parameters and higher order interactions in the inflaton potential. Then, we consider the Jordan-frame Lagrangian for the pole  inflation for the SM Higgs doublet $H$ \cite{Clery:2023ptm} with a conformal coupling and a minimal kinetic term, as follows,
\bea
\frac{{\cal L}_J}{\sqrt{-g_J}} = -\frac{1}{2}M^2_P\, \Omega(H) R(g_J) + |D_\mu H|^2 -V_J(H) \label{LJ}
\eea
where the non-minimal coupling function is given by 
\bea
\Omega(H) &=& 1-\frac{1}{3M^2_P}|H|^2.
\eea
For the PQ inflation \cite{Lee:2023dtw,Lee:2024bij}, we only have to replace the SM doublet $H$ by the PQ complex field $\Phi$ and the covariant derivative becomes the usual derivative.

Then, after making a Weyl transformation of the metric by $g_{J,\mu\nu}=g_{E,\mu\nu}/\Omega$ with $\Omega=1-\frac{1}{3M^2_P}|H|^2$, we obtain the Einstein-frame Lagrangian as follows,
\bea
\frac{{\cal L}_E}{\sqrt{-g_E}} &=&-\frac{1}{2} M^2_P R(g_E) + \frac{|D_\mu H|^2}{\big(1-\frac{1}{3M^2_P}|H|^2\big)^2} \nonumber \\
&&-\frac{1}{3M^2_P\big(1-\frac{1}{3M^2_P}|H|^2\big)^2}\bigg(|H|^2 |D_\mu H|^2-\frac{1}{4}\partial_\mu |H|^2 \partial^\mu |H|^2\bigg) -V_E(H) \label{LE}
\eea
where the Einstein-frame Higgs potential becomes
\bea
V_E(H)= \frac{V_J(H)}{\big(1-\frac{1}{3M^2_P}|H|^2\big)^2}.
\eea
For instance, we take a simple form of the Jordan-frame Higgs potential, as follows,
\bea
V_J(H) = c_m \Lambda^{4-2m} |H|^{2m} \bigg(1-\frac{1}{3M^2_P}|H|^2\bigg)^2,  \label{poleJ}
\eea
for which the Einstein-frame potential becomes
\bea
V_E(H) = c_m \Lambda^{4-2m} |H|^{2m}. 
\eea
In the following discussion, we take $M_P=1$, but we recover $M_P$ for the formulas if more clear. 

We note that the form of the Jordan-frame Higgs potential in eq.~(\ref{poleJ}) is motivated from the Weyl invariant Lagrangian of the Higgs inflation \cite{Lee:2024rjw}.  In this case, the form of the Higgs potential is related to the form of the non-minimal coupling to gravity in the limit of the non-compact isometry $SO(1,N)$ in the field space for the dilaton $\chi$ and the Higgs doublet $H$ with $N=4$ (or the PQ field with $N=2$). For instance, the Weyl invariant Lagrangian for the Higgs doublet with an approximate $SO(1,4)$ isometry  \cite{Lee:2024rjw} is given by
\bea
\frac{{\cal L}_J}{\sqrt{-g_J}}&=&(1+a)\bigg[-\frac{1}{12}  (\chi^2-2|H|^2 ) R-\frac{1}{2}(\partial_\mu\chi)^2+|D_\mu H|^2 \bigg]  \nonumber \\
&&+ \frac{1}{2}a({\hat D}_\mu\chi)^2 - a |{\hat D}_\mu H|^2 -\frac{1}{4} w_{\mu\nu}w^{\mu\nu} - V_J(\chi, H)\label{Lag}
\eea
where $w_\mu$ is the Weyl gauge field, $a$ is a constant parameter, ${\hat D}_\mu\chi=(\partial_\mu-g_w w_\mu)\chi$,  ${\hat D}_\mu H$ contains the covariant derivative containing Weyl gauge field by ${\hat D}_\mu H=(\partial_\mu-g_w w_\mu+\cdots)H$, and $D_\mu H$ is the covariant derivative in the SM. Here, the Jordan frame potential takes
\bea
V_J(\chi, H)=\frac{1}{\langle\chi^4\rangle}\, f(|H|^2/\chi^2) (\chi^2-2 |H|^2)^2. \label{Weylpot}
\eea
Here, $f(|H|^2/\chi^2) $ is an arbitrary function preserving the Weyl invariance but it generically breaks the $SO(1,4)$ isometry explicitly. If $f(|H|^2/\chi^2) $ is constant, the above Jordan frame potential is a unique choice with renormalizability and Weyl invariance. However, for a general non-constant  $f(|H|^2/\chi^2) $ keeps only the  $SO(4)$ subgroup, the pole inflation takes place for an appropriate choice of the parameters. In this case, we get the relation between the Jordan frame potential and the frame function by $V_J\propto \Omega(H)^2 $.  For the Higgs pole inflation \cite{Lee:2024rjw}, we choose 
\bea
f(|H|^2/\chi^2)=m^2_H \langle \chi^2\rangle\,\cdot \frac{|H|^2}{\chi^2}+\lambda_H \langle\chi^4\rangle\,\cdot \frac{|H|^4}{\chi^4}.
\eea
After the gauge fixing of the Weyl symmetry with $\langle\chi\rangle=\sqrt{6/(1+a)}$, the small violation of the non-compact isometry leads to the Einstein frame potential, $V_E=V_J/\Omega^2=\lambda_H |H|^4$. Moreover, the Weyl gauge field  couples to gravity minimally and it gets mass, $m^2_w=6a g^2_w/(1+a)$ \cite{Lee:2024rjw}, after the gauge fixing.  The extra couplings due to the Weyl covariant derivatives do not change the inflationary predictions of the Higgs pole inflation much for $|a|\ll 1$ \cite{Lee:2024rjw}.
A similar discussion can be applied to the case of the PQ pole inflation.

We also note that other $SO(1,4)$ breaking parameters could be also introduced in the Weyl invariant potential, for instance, $(\chi^2-2\kappa |H|^2)^2$, with $\kappa\neq 1$, instead of $ (\chi^2-2 |H|^2)^2$ in eq.~(\ref{Weylpot}). In this case, a deviation from the pole inflation must be limited for a slow-roll inflation, namely, $|\kappa-1|\ll 1$. We postpone the detailed analysis of more general $SO(1,4)$ breaking effects to a future work.

For breaking the electroweak symmetry or the PQ symmetry spontaneously, we need to introduce a tachyonic mass term in the Einstein frame, $V_E\supset -m^2_H |H|^2$, but it is not important for inflation.

\section{RG-improved inflaton potential}

We consider the Lagrangian for the Higgs pole inflation in the Einstein frame at both tree and one-loop levels and compare it with the case in the PQ pole inflation.

\subsection{Higgs Lagrangian in Einstein frame}

In the unitary gauge where the Higgs doublet is taken to $H^T=(0,h)^T/\sqrt{2}$, the Higgs kinetic terms in the first two terms in the second line of eq.~(\ref{LE}) are cancelled. Then, we get the Einstein-frame Lagrangian for the Higgs boson $h$ as
\bea
\frac{{\cal L}_E}{\sqrt{-g_E}} =-\frac{1}{2} M^2_P R +\frac{1}{2}\,\frac{(\partial_\mu h)^2}{\big(1-\frac{1}{6M^2_P}h^2\big)^2} - \frac{V_J\big(\frac{1}{\sqrt{2}}h\big)}{\big(1-\frac{1}{6M^2_P}h^2\big)^2}. \label{unit}
\eea
Then, making the Higgs kinetic term canonically normalized by
\bea
h=\sqrt{6}M_P \tanh\Big(\frac{\phi}{\sqrt{6}M_P}\Big), \label{can}
\eea
we rewrite the Einstein-frame Lagrangian in eq.~(\ref{unit}) for the Higgs potential in eq.~(\ref{poleJ}) as
\bea
\frac{{\cal L}_E}{\sqrt{-g_E}} =-\frac{1}{2} M^2_P R + \frac{1}{2}(\partial_\mu\phi)^2 - V_E(\phi),
\eea
with 
\bea
V_E(\phi)= 3^m c_m\Lambda^{4-2m}M^{2m}_P \bigg[  \tanh\Big(\frac{\phi}{\sqrt{6}M_P}\Big)\bigg]^{2m}. \label{inflatonpot}
\eea 
As a result, a slow-roll inflation is possible at large $|\phi|$ or $|h|\sim \sqrt{6} M_P$ near the pole of the general Higgs kinetic term in eq.~(\ref{unit}).
For the Higgs quartic potential with $m=2$, we take $c_2=\lambda_H$, namely, $V_E=\lambda_H |H|^4=\frac{1}{4}\lambda_H h^4$, for which the inflaton potential in eq.~(\ref{inflatonpot}) becomes
\bea
V_E(\phi)=9\lambda_H M^4_P\bigg[  \tanh\Big(\frac{\phi}{\sqrt{6}M_P}\Big)\bigg]^{4}. \label{tree}
\eea

\subsection{Coleman-Weinberg potential for the Higgs inflaton}

Using the chain rules for derivatives, 
\bea
\frac{dV_E}{d\phi}&=& \frac{dh}{d\phi}\cdot \frac{d V_E}{dh}=\Omega\,\frac{dV_E}{d h}, \\
\frac{d^2V_E}{d\phi^2}&=& \frac{d\Omega}{d\phi}\cdot \frac{dV_E}{dh} +\Omega^2 \frac{d^2V_E}{dh^2} \nonumber \\
&=&-\frac{1}{3}\Omega \lambda_H h^4 +3\Omega^2 \lambda_H h^2,
\eea
with $\Omega=1-\frac{1}{6}h^2$,
we obtain the effective Higgs mass during inflation as
\bea
M^2_\phi&=&\frac{d^2 V_E}{d^2\phi} \nonumber \\
&=& 3\lambda_H h^2 \Big(1-\frac{1}{6}h^2\Big) \Big(1-\frac{5}{18}h^2\Big).
\eea
Thus. the effective Higgs mass gets suppressed as compared to the Hubble expansion rate during inflation, satisfying the slow-roll condition,  $|\eta|\ll 1$, with
\bea
\eta&=&\frac{M^2_P M^2_\phi}{V_E} \nonumber \\
&\simeq& -\frac{8M^2_P}{h^2}\Big(1-\frac{1}{6}h^2\Big) \simeq -\frac{16}{3}\, e^{-2\phi/\sqrt{6}}.
\eea

From eq.~(\ref{LE}), we also obtain the effective masses for the electroweak gauge bosons as 
\bea
M^2_W &=& \frac{g^2h^2}{4(1-\frac{1}{6}h^2)}=\frac{3}{2}g^2M^2_P \sinh^2\Big(\frac{\phi}{\sqrt{6}M_P}\Big), \\
M^2_Z &=& \frac{(g^2+g^{\prime 2})h^2}{4(1-\frac{1}{6}h^2)}=\frac{3}{2}(g^2+g^{\prime 2})M^2_P \sinh^2\Big(\frac{\phi}{\sqrt{6}M_P}\Big).
\eea
Thus, during inflation with $\phi\gg \sqrt{6}M_P$, $M^2_W\gg \frac{2}{3}g^2M^2_P$ and $M^2_Z\gg \frac{3}{2}(g^2+g^{\prime 2})M^2_P$, so they are decoupled from the inflaton. On the other hand, after inflation, $\phi\ll \sqrt{6}M_P$, for which $M^2_W\simeq \frac{1}{4}g^2 \phi^2 $ and $M^2_Z\simeq \frac{1}{4}(g^2+g^{\prime 2})\phi^2$, so we can recover the standard interactions of the Higgs boson to the electroweak gauge bosons, so reheating can proceed.  

In the $R_\xi$ gauge, such as Landau gauge, we also need to consider the Goldstone contributions to the CW potential. 
Writing the Higgs doublet as $H^T=\frac{1}{\sqrt{2}}(G_1+iG_2,h+iG_3)$, the quadratic terms for the Goldstone bosons, $G_i (i=1,2,3)$, are given by
\bea
\frac{{\cal L}_E}{\sqrt{-g_E}}\supset \sum_{i=1}^3\left[\frac{1}{2}\,\frac{(\partial_\mu G_i)^2}{(1-\frac{1}{6}h^2)}- \frac{1}{2}\lambda_H h^2 G^2_i \right],
\eea
while the kinetic mixing terms between the inflaton $h$ and the Goldstones can be ignored. Thus, in Landau gauge, after the kinetic terms for the Goldstone bosons are canonically normalized, the effective masses for  for the Goldstone bosons are given by
\bea
m^2_{G_i}=\lambda_H h^2 \Big(1-\frac{1}{6}h^2\Big), \quad i=1,2,3,
\eea
which are suppressed for $h\to \sqrt{6}$ during inflation. 

We also consider the Yukawa interactions between the Higgs and the SM fermions in the Jordan frame,
\bea
\frac{{\cal L}_{J,{\rm Yukawa}}}{\sqrt{-g_J}}=-\frac{1}{\sqrt{2}} y_f h {\bar f} f.
\eea
Then, after the Weyl rescaling of the metric with $g_{J,\mu\nu}=g_{E,\mu\nu}/\Omega$ and the redefinition of the fermions by $f'=\Omega^{-3/4}f$ \cite{Weyl}, we obtain the effective Yukawa interactions in the Einstein frame as
\bea
\frac{{\cal L}_{E,{\rm Yukawa}}}{\sqrt{-g_E}}=-\frac{1}{\sqrt{2}} \Omega^{-1/2} y_f h {\bar f}' f'.
\eea
Therefore, the effective fermion masses in the SM are given by
\bea
m_f=\frac{1}{\sqrt{2}} \Omega^{-1/2} y_f h =\frac{1}{\sqrt{2}} \,\frac{y_f h}{\sqrt{1-\frac{1}{6}h^2}}.
\eea
As a result, we find that $W, Z$ and fermion masses are proportional to $h/\sqrt{\Omega}$, whereas the Higgs and Goldstone masses are proportional to $h\sqrt{\Omega}$, which are suppressed during inflation.

After summing up the Higgs interactions to the SM, we get the one-loop renormalized Coleman-Weinberg potential for the inflaton in Landau gauge by
\bea
V_{\rm CW}= \sum_\alpha \frac{N_\alpha}{64\pi^2} M^4_\alpha \bigg(\ln \frac{M^2_\alpha}{\mu^2}-C_\alpha\bigg)
\eea
where $\alpha=\{Z,W,t,h,G\}$ for $Z, W$ gauge bosons, top quark, Higgs and Goldstones, respectively, with $N_\alpha=\{3,6, -12, 1, 3\}$, and $C_\alpha=\frac{3}{2}$ for fermions or scalars and $C_\alpha=\frac{5}{6}$ for gauge bosons.
For large Higgs field values during inflation, we choose the renormalization scale to $\mu= c_*h/\sqrt{\Omega}$ with $c_*$ being the optimal value of order one in order to keep the logarithms minimized in the perturbative expansion of the Coleman-Weinberg potential \cite{RG,RG2,Casas:1994qy} as the Higgs field value changes. One could choose another renormalization scale, for instance, a field-independent constant value. In this case, the CW potential is  field-dependent and dominated by $\ln (h/\sqrt{\Omega})$, while $\lambda(\mu)$ in the tree-level potential is constant, leading to the same results as in the case with $\mu= c_*h/\sqrt{\Omega}$.  Then, the results do not depend on the choice of the renormalization scale in the full theory \cite{RG2}. 
In our work, we make a choice for the renormalization scale such that the logarithms in the one-loop Coleman-Weinberg potential are minimized and the loop effects in the effective potential are encoded in the running couplings.

Following the procedure described in the Appendix A, we find that the renormalized effective potential during inflation is expanded for $h^2<6$ as
\bea
V_E(h)\approx \frac{1}{4} \lambda_H(h) h^4+\frac{1}{6} \lambda_6(h) h^6
\eea
where the running quartic coupling $\lambda_H$ becomes field-dependent due to  $\mu=c_*h/\sqrt{\Omega}$ and we introduced the sextet coupling $\lambda_6$ for the Higgs after the renormalization of the CW potential\footnote{See the details for the renormalization of the CW potential at one-loops in the Appendix A.}. Henceforth, as will be shown shortly, choosing the renormalized sextet coupling and the higher order self-couplings for the Higgs to small values by the renormalization conditions such that they are sub-dominant for inflation, we focus on the effects of the running Higgs quartic coupling.
We note that the Higgs and Goldstone contributions to the CW potential can be suppressed, relative to $W, Z$ boson and top quark contributions, but they are important for deriving the RG equations as shown in the Appendix A.

\begin{figure}[!t]
\begin{center}
 \includegraphics[width=0.42\textwidth,clip]{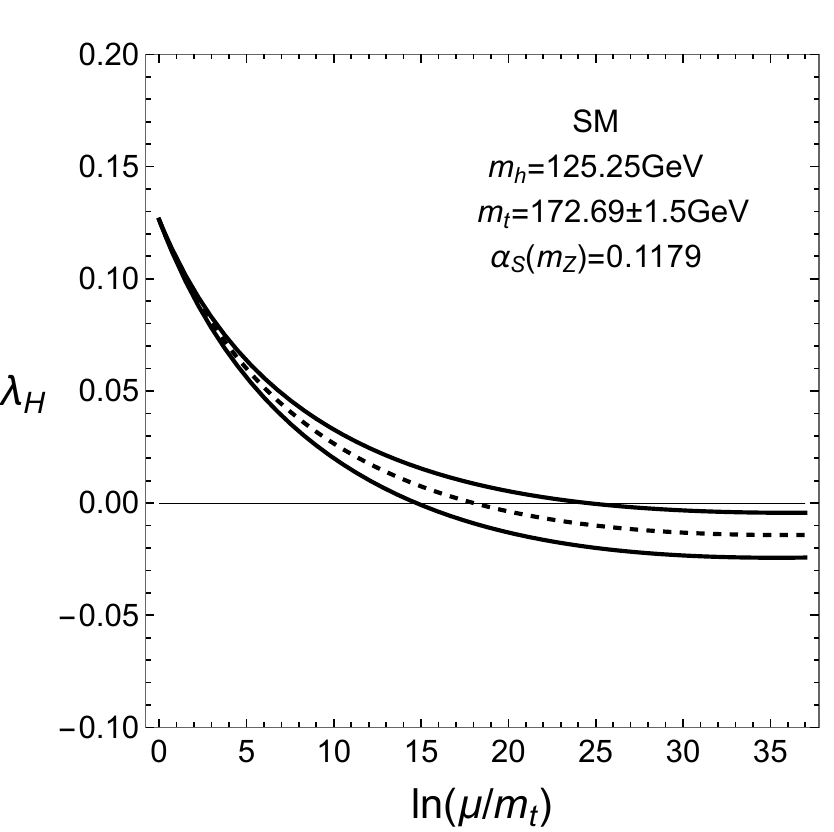}
 \end{center}
\caption{The full two-loop running Higgs quartic coupling $\lambda_H$ in the SM. We took the top quark mass to $m_t=172.69\pm 1.5\,{\rm GeV}$ along the solid lines and $m_t=172.69\,{\rm GeV}$ along the dashed line. }
\label{fig:SM}
\end{figure}

We also note that there can be extra contributions to the Coleman-Weinberg potential due to new singlet scalar fields. For instance, introducing a Higgs-portal coupling to a real singlet scalar $S$ in the Jordan frame Lagrangian \cite{Clery:2023ptm} by
\bea
\frac{{\cal L}_{J,S}}{\sqrt{-g_J}} = \frac{1}{2}(\partial_\mu S)^2 -\frac{1}{2}m^2_S S^2-\frac{1}{4}\lambda_S S^4 - \lambda_{HS} S^2 |H|^2,
\eea
we find that the Einstein-frame Lagrangian for the singlet scalar becomes
\bea
\frac{{\cal L}_{E,S}}{\sqrt{-g_E}} =  \frac{1}{2\Omega}(\partial_\mu S)^2-\bigg(\frac{1}{2}m^2_S S^2+\frac{1}{4}\lambda_S S^4+ \lambda_{HS} S^2 |H|^2\Big)\Omega^{-2}.
\eea
Thus, taking into account the rescaling of the singlet scalar by $S\to \sqrt{\Omega}\,S$, we obtain the effective singlet scalar mass for a slowly varying inflaton as
\bea
M^2_S=m^2_S+\frac{\lambda_{HS} h^2}{\Omega}. \label{smass}
\eea
Therefore, for $\lambda_{HS} h^2/\Omega \gg m^2_S$,  the singlet scalar contributes to the one-loop CW potential with the same dependence on the Higgs field (i.e. $h/\sqrt{\Omega}$) as for $W, Z$ gauge bosons and SM fermions. Moreover, when the singlet scalar field gets a VEV and it is heavier than the Higgs boson, the Higgs mixing gives rise to a tree-level shift in the running Higgs quartic coupling \cite{VSB} by
\bea
\lambda_H=\lambda_{H,{\rm SM}} + \frac{\lambda^2_{HS}}{\lambda_S}, \label{shift}
\eea
so the Higgs quartic coupling can remain positive at scales much larger than the one in the SM such as the inflation scale.

From eq.~(\ref{smass}), we remark that the effective mass for the singlet scalar becomes large during inflation with $h\to \sqrt{6}$ due to the Higgs portal coupling.  In comparison to the Hubble scale during inflation, which is given by $H^2_I\simeq 3\lambda_H M^2_P$, the singlet scalar is safely decoupled for $M^2_S\gg H^2_I$, which becomes $2\lambda_{HS}\sinh^2(\phi/(\sqrt{6}M_P))\gg \lambda_H$ during inflation. Thus, for $\phi\gg M_P$, we only need a non-negligible $\lambda_{HS}$, which is similar to or larger than $\lambda_H$ during inflation. In this case, the inflaton dynamics is dominated by the Higgs boson, although there are effects on the inflaton potential through the loop corrections of the singlet scalar, as will be shown later.

We draw the running Higgs quartic coupling in Fig.~\ref{fig:SM} within the SM, showing that it can be positive up to $\sim 10^{10}\,{\rm GeV}$ around the inflation scale \cite{Elias-Miro:2011sqh}. Although the top quark pole mass is still subject to large experimental uncertainties, it is plausible that new physics contributions could alter the beta function for the Higgs quartic coupling such that the Higgs quartic coupling is small and positive around the inflation scale as far as the new physics scale appears below the instability scale of the SM vacuum \cite{Clery:2023ptm}.

\begin{figure}[!t]
\begin{center}
 \includegraphics[width=0.42\textwidth,clip]{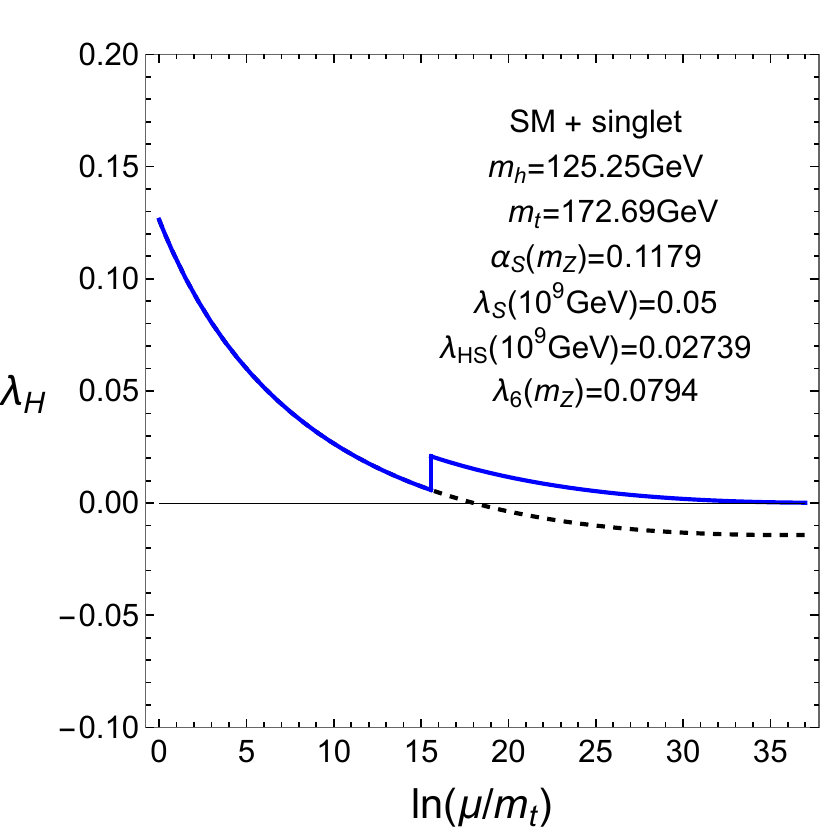}\,\, \,\, \includegraphics[width=0.45\textwidth,clip]{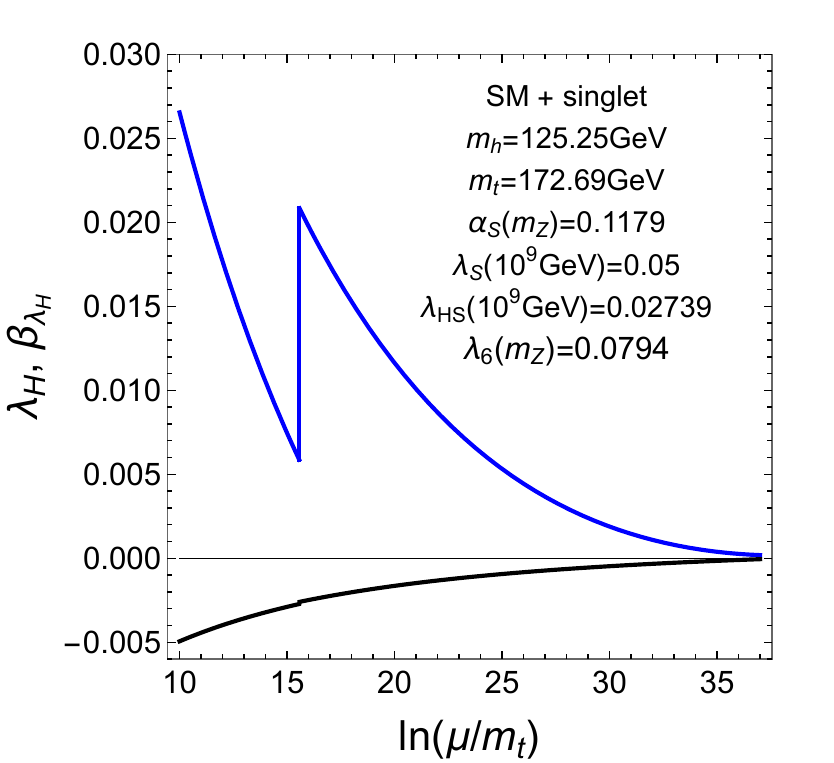}  \\
  \includegraphics[width=0.42\textwidth,clip]{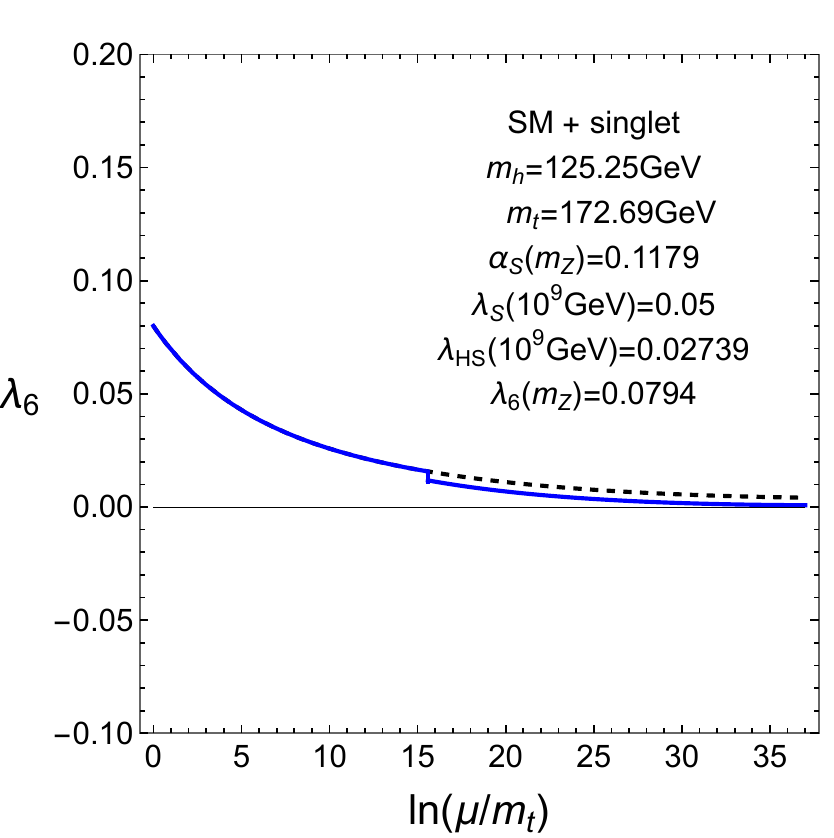}\,\, \,\, \includegraphics[width=0.45\textwidth,clip]{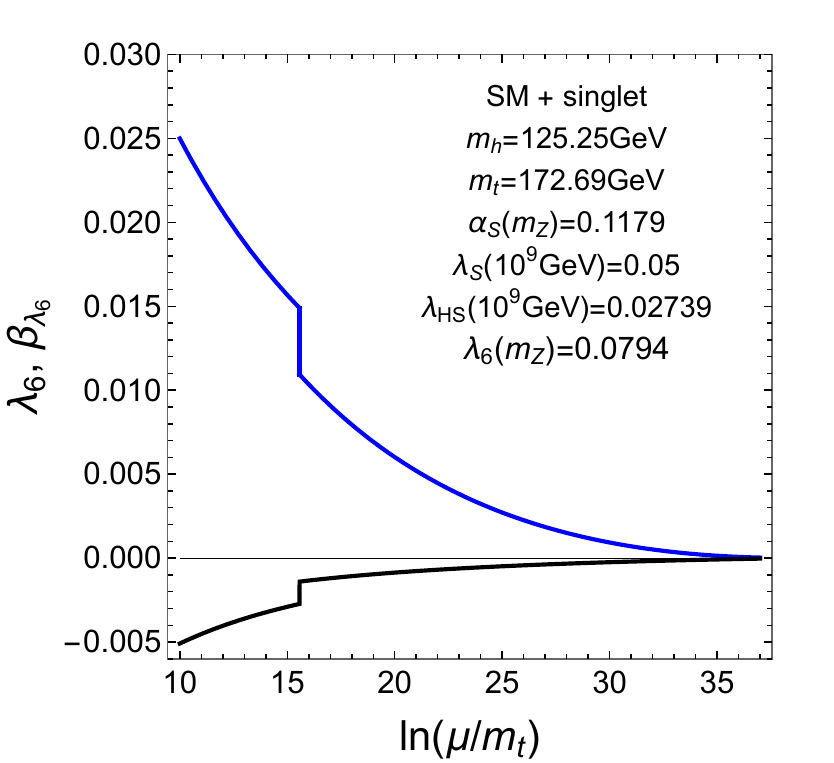} 
 \end{center}
\caption{Upper: The full two-loop running Higgs quartic coupling $\lambda_H$ for the SM  in black dashed line and for the SM plus the singlet scalar in blue solid line (Left). The running  $\lambda_H$ and its beta function in blue and black lines, respectively (Right).  The beta function remains negative all the way to $\mu\sim M_P$. Lower: The one-loop running sextet coupling $\lambda_6$ for the SM  in black dashed line and for the SM plus the singlet scalar in blue solid line (Left). The running Higgs sextet coupling $\lambda_6$ in units of $M^{-2}_P$ and its beta function in blue and black lines, respectively (Right).  The one-loop running sextet coupling remains smaller than the Higgs quartic coupling.}
\label{fig:RG1}
\end{figure}

\begin{figure}[!t]
\begin{center}
 \includegraphics[width=0.42\textwidth,clip]{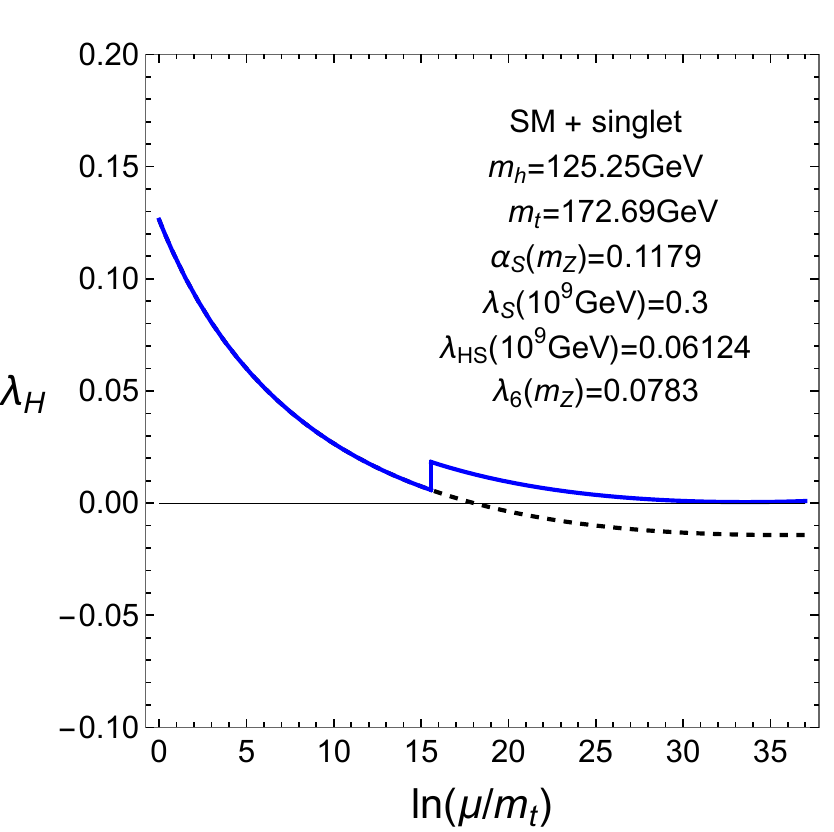}\,\, \,\, \includegraphics[width=0.45\textwidth,clip]{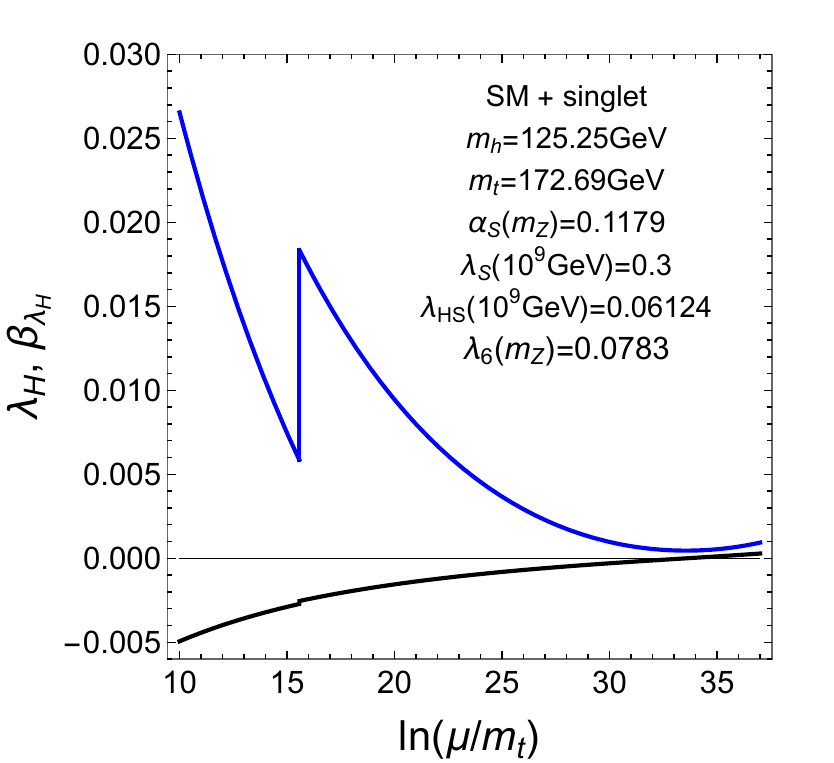}  \\
  \includegraphics[width=0.42\textwidth,clip]{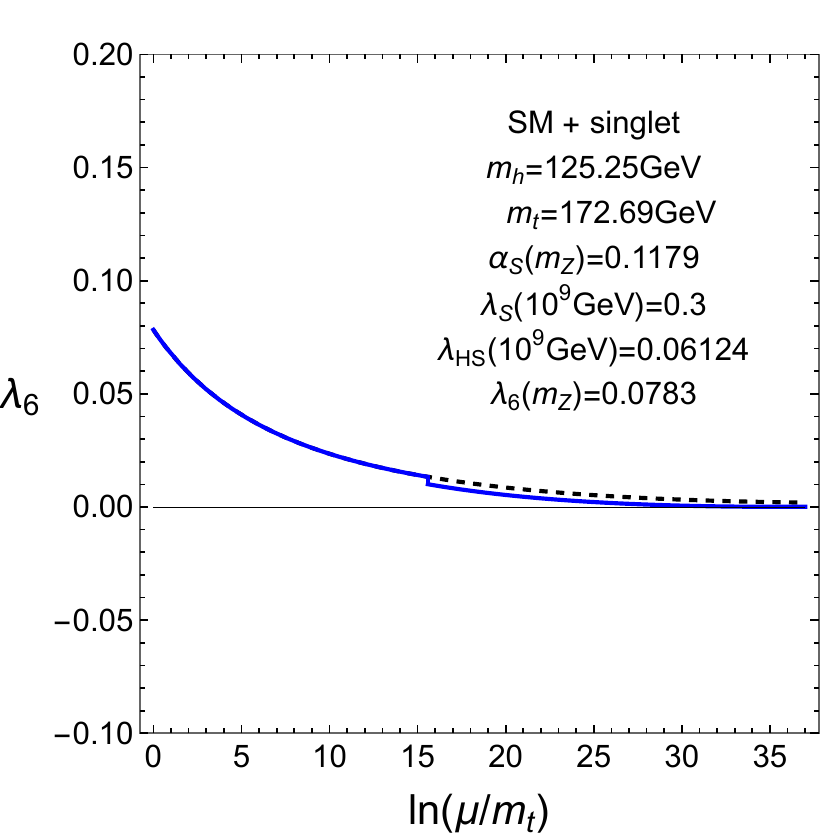}\,\, \,\, \includegraphics[width=0.45\textwidth,clip]{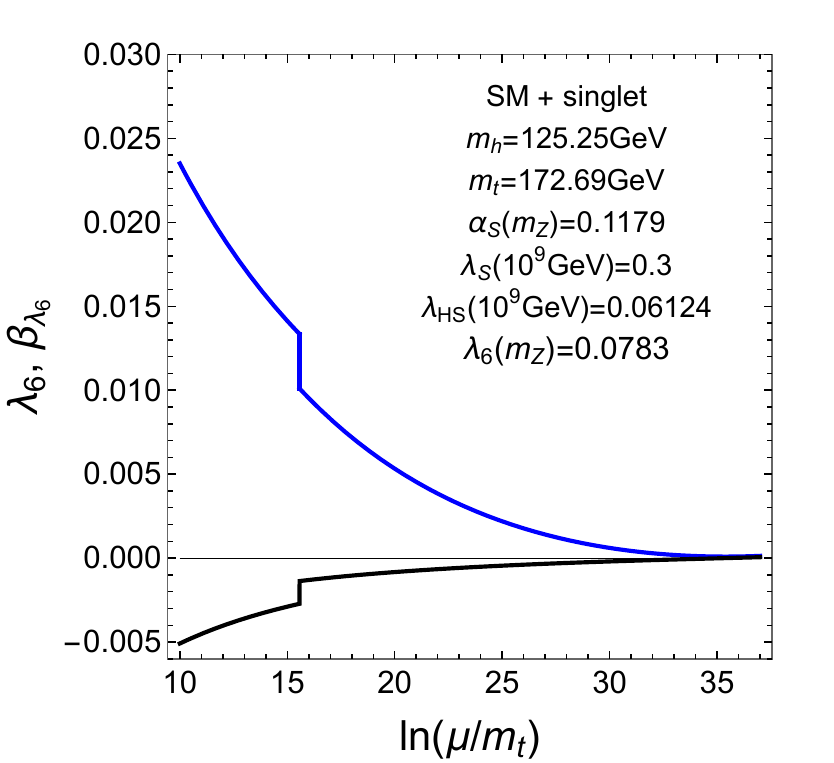} 
 \end{center}
\caption{The same as in Fig.~\ref{fig:RG1}, except for relatively larger couplings for the singlet scalar. The beta function for the Higgs quartic coupling becomes a positive value near $\mu\sim M_P$. }
\label{fig:RG2}
\end{figure}

In the upper panel of Fig.~\ref{fig:RG1}, we depict the running Higgs quartic coupling from the two-loop renormalization group equations \cite{RG,Das:2015nwk}  in blue lines, after the tree-level shift and loop corrections due to the singlet scalar with mass $m_S=10^9\,{\rm GeV}$ are included on both sides and  the beta function for the Higgs quartic coupling is shown in black line in the right plot.  We chose   $\lambda_S=0.05$ and  $\lambda_{HS}=0.02739$ at the singlet scalar threshold. On the other hand, the running Higgs quartic coupling in the SM  is shown in dashed black line in the left plot. We find that in the presence of the singlet scalar couplings, the Higgs quartic coupling is positive and small while  the beta function remains negative and small toward  $\mu\sim M_P$.  

In comparison, in the lower panel of Fig.~\ref{fig:RG1}, we also show the running Higgs sextet coupling and its beta function at one-loop level, for the same set of the parameters for the singlet scalar and $\lambda_6=0.0794$ at the $Z$-pole in units of $M^{-2}_P$. As a result, we find that the running sextet potential remains smaller than the Higgs quartic potential during inflation for an appropriate sextet coupling at low energy. Similar results would apply to the running couplings of the higher order terms in the Higgs potential,  justifying the dominance of the Higgs quartic potential for inflation.

In the upper panel of Fig.~\ref{fig:RG2}, we also show the case for $\lambda_S=0.3$ and $\lambda_{HS}=0.06124$ at the singlet threshold with $m_S=10^9\,{\rm GeV}$. In this case, the Higgs quartic coupling gets positive and small near the Planck scale and the beta function for the Higgs quartic coupling also turns positive and small. In the lower panel of Fig.~\ref{fig:RG2}, we obtain the running Higgs sextet coupling for the same set of the parameters for the singlet scalar and $\lambda_6=0.0783$ at the $Z$-pole in units of $M^{-2}_P$. Therefore, we draw a similar conclusion as in Fig.~\ref{fig:RG1} that the running sextet potential can remain smaller than the Higgs quartic potential during inflation.

Assuming that new physics contributions to the running of the Higgs quartic couplings exists in terms of the tree-level shift in eq.~(\ref{shift}) and the loop corrections in the Coleman-Weinberg potential with eq.~(\ref{smass}), we parametrize the running Higgs quartic coupling as being running up to two-loops near the inflation scale $\mu_I$, as follows, 
\bea
\lambda_H(\mu)=\lambda(\mu_I) + b_1 \bigg(\ln \frac{\mu}{\mu_I}\bigg)+ b_2 \bigg(\ln \frac{\mu}{\mu_I}\bigg)^2, \label{twoloops}
\eea
with $\mu=c_*h/\sqrt{\Omega}$ and $b_{1,2}$ being the one-loop and two-loop beta function coefficients, respectively.
We treat $b_{1,2}$ to be model-dependent parameters encoding the Higgs couplings of the extra particles near the inflation scale and assume that the running Higgs quartic coupling remains small and positive during inflation.

\subsection{The case for the PQ inflaton}

We also comment on the possibility that the inflaton is identified as the radial component of the PQ field.
In this case, taking $\Phi=\frac{1}{\sqrt{2}}\rho\,e^{\i\theta}$, the Einstein frame Lagrangian for the PQ field \cite{Lee:2023dtw,Lee:2024bij} becomes
\bea
\frac{{\cal L}_E}{\sqrt{-g_E}} =-\frac{1}{2} M^2_P R +\frac{1}{2}\,\frac{(\partial_\mu \rho)^2}{\big(1-\frac{1}{6M^2_P}\rho^2\big)^2} +\frac{1}{2}\frac{\rho^2(\partial_\mu\theta)^2}{\big(1-\frac{1}{6M^2_P}\rho^2\big)}- \frac{V_J\big(\frac{1}{\sqrt{2}}\rho\big)}{\big(1-\frac{1}{6M^2_P}\rho^2\big)^2}. 
\eea 
Then, making the radial mode kinetic term canonically normalized by
\bea
\rho=\sqrt{6}M_P \tanh\Big(\frac{\phi}{\sqrt{6}M_P}\Big) \label{can2}
\eea
and following the similar steps as in Section 3.1, we reach the Einstein-frame Lagrangian in terms of the canonical inflaton, as follows,
\bea
\frac{{\cal L}_E}{\sqrt{-g_E}} =-\frac{1}{2} M^2_P R + \frac{1}{2}(\partial_\mu\phi)^2+3\sinh^2\Big(\frac{\phi}{\sqrt{6}M_P}\Big) (\partial_\mu\theta)^2 - V_E(\phi).
\eea
For $V_E=\lambda_\Phi |\Phi|^4=\frac{1}{4}\lambda_\Phi \rho^4$, the inflaton potential becomes
\bea
V_E(\phi)=9\lambda_\Phi M^4_P\bigg[  \tanh\Big(\frac{\phi}{\sqrt{6}M_P}\Big)\bigg]^{4}. \label{tree2}
\eea
We note that the angular component of the PQ field or the axion $\theta$ is massless before the QCD phase transition or if an explicit breaking of the PQ symmetry is ignored. However, there is no problem with the axion isocurvature perturbations due to a large effective axion decay constant during inflation \cite{Lee:2024bij,Lee:2024rjw}.

In KSVZ models \cite{KSVZ}, the PQ field couples to a heavy vector-like quark $Q$, the SM Higgs and right-handed neutrinos $N_R$ with appropriate PQ charges \cite{Lee:2023dtw}, in Jordan frame, as follows,
\bea
\frac{{\cal L}_{J,{\rm KSVZ}}}{\sqrt{-g_J}}= \bigg(-y_Q \Phi\,  \overline {Q_L} Q_R -\frac{1}{2}y_N \Phi \,\overline{N^c_R}N_R +{\rm h.c.}\bigg)-2\lambda_{H\Phi}|\Phi|^2|H|^2.
\eea 
On the other hand, in DFSZ models \cite{DFSZ}, the PQ field couples to two Higgs doublets, $H_1$ and $H_2$, and right-handed neutrinos $N_R$ with appropriate PQ charges\cite{Lee:2024bij}, in Jordan frame, as follows,
\bea
\frac{{\cal L}_{J,{\rm DFSZ}}}{\sqrt{-g_J}}=\bigg( -\frac{1}{2}y_N \Phi\, \overline{N^c_R}N_R -\kappa \Phi^2 H^\dagger_1 H_2 +{\rm h.c.}\bigg) - 2\lambda_{1\Phi}|\Phi|^2|H_1|^2- 2\lambda_{2\Phi}|\Phi|^2|H_2|^2.
\eea
In these models, there are model-dependent interactions between two Higgs doublets  and the SM particles \cite{Lee:2024bij}.

In both KSVZ and DFSZ models, the effective mass for the inflaton is given by 
\bea
M^2_\phi=3\lambda_\Phi\rho^2  \Big(1-\frac{1}{6}\rho^2\Big) \Big(1-\frac{5}{18}\rho^2\Big),
\eea
which is suppressed for $\rho\to \sqrt{6}$ during inflation. Similarly as in the fermion masses in the Higgs pole inflation, the masses for the heavy vector-like quark in KSVZ models  and the right-handed neutrinos in both models are given by
\bea
M_Q &=& \frac{1}{\sqrt{2}}\, \frac{y_Q \rho}{\sqrt{1-\frac{1}{6}\rho^2}}, \\
M_N &=& \frac{1}{\sqrt{2}}\, \frac{y_N \rho}{\sqrt{1-\frac{1}{6}\rho^2}}.
\eea 
The Higgs portal couplings give rise to the inflaton-dependent effective masses for the Higgs fields, as follows,
\bea
M^2_H &=& \frac{\lambda_{H\Phi}\rho^2}{1-\frac{1}{6}\rho^2}, \quad {\rm KVSZ}, \\
M^2_{{\tilde H}_1,{\tilde H}_2} &=&  \frac{1}{2}\Big[\lambda_{1\Phi}+\lambda_{2\Phi}\pm \sqrt{(\lambda_{1\Phi}-\lambda_{2\Phi})^2+\kappa^2}  \Big] \frac{\rho^2}{1-\frac{1}{6}\rho^2}, \quad {\rm DFSZ}
\eea
where the Higgs fields in DFSZ models are redefined to ${\tilde H}_1,{\tilde H}_2$ for the diagonalized mass matrix.

As a result, as in the Higgs pole inflation, the one-loop CW potential for the inflaton is given by
\bea
V_{\rm CW}= \sum_\alpha \frac{N_\alpha}{64\pi^2} M^4_\alpha \bigg(\ln \frac{M^2_\alpha}{\mu^2}-C_\alpha\bigg)
\eea
where $\alpha=\{\rho,H,Q,N\}$ with $N_\alpha=\{1, 4,-12, -6\}$ in KSVZ models and  $\alpha=\{\rho,{\tilde H}_1,{\tilde H}_2,N\}$ with $N_\alpha=\{1, 4,4, -6\}$ in DFSZ models,
In both KSVZ and DFSZ models, the effective masses for the fields coupled to the inflaton are proportional to $\rho/\sqrt{\Omega(\rho)}$ with $\Omega(\rho)=1-\frac{1}{6}\rho^2$, other than the inflaton mass. Then, after following the renormalization procedure in the Appendix A and choosing the optimal value for the renormalization scale by $\mu=d_* \rho/\sqrt{\Omega(\rho)}$ with $d_*$ being of order one, we expand the effective potential for $\rho^2<6$ by
\bea
V_E(\rho)\approx \frac{1}{4} \lambda_\Phi(\rho) \rho^4+\frac{1}{6} {\bar\lambda}_6(\rho) \rho^6,
\eea
with the field-dependent quartic coupling $\lambda_\Phi(\rho)$ and the sextet coupling ${\bar\lambda}_6$ introduced after the renormalization of the CW potential. Similarly to the Higgs pole inflation, we choose the renormalization conditions for the renormalized sextet coupling and the higher order self-couplings for the PQ field by setting them to small values at the inflation scale to be sub-dominant during inflation, so we focus only on the running quartic coupling for the PQ field up to two-loops as in eq.~(\ref{twoloops}) in the previous section. But, the beta function coefficients, $b_1, b_2$, in eq.~(\ref{twoloops}), are model-dependent, and they can be different from those in the Higgs pole inflation.

For the PQ pole inflation, the beta function coefficients of the Higgs quartic coupling also depend on $y_Q, y_N., \lambda_{H\Phi}$ in KSVZ models and $y_N, \kappa, \lambda_{1\Phi}, \lambda_{2\Phi}$ in DFSZ models, in contrast to  the Higgs pole inflation where a singlet scalar field can be added to improve the vacuum instability. Thus, in this case, a small running quartic coupling for the PQ field is maintained in the presence of small couplings of the PQ fields, namely, as far as $y_Q, y_N\lesssim 10^{-3}$ and $\lambda_{H\Phi}, \kappa, \lambda_{1\Phi}, \lambda_{2\Phi}\lesssim 10^{-5}$ \cite{Lee:2023dtw,Lee:2024bij}, and the beta function can be chosen to be small and of either signs for appropriate choices of new couplings. Therefore, we don't pursue a concrete analysis of the running quartic coupling for the PQ field.

We remark on the model discrimination in light of the ACT results. We find that the model parameter space favored by the ACT results can be inferred only in terms of the beta functions of the quartic couplings in both Higgs and PQ inflation models, which depend on various couplings of the inflaton to the SM and extra fields. However, unlike the Higgs pole inflation where the Higgs inflaton has sizable couplings to the SM and the singlet scalar field, we can take the PQ inflaton couplings to be small enough throughout the RG evolution such that a small quartic coupling for the PQ field during inflation can be obtained for the CMB normalization.  The reheating process after inflation is efficient for Higgs pole inflation \cite{Clery:2023ptm}, but it can be ineffective in the case for PQ pole inflation \cite{Lee:2023dtw,Lee:2024bij}. Therefore, we could distinguish between Higgs and PQ pole inflation models by the post-inflationary evolution of the inflaton and the reheating temperature \cite{Clery:2023ptm,Lee:2023dtw,Lee:2024bij,Lee:2024rjw}.

\section{Pole inflation with loops}

We discuss the parametrization of loop corrections to the inflaton potential in the pole inflation up to two-loop orders and discuss the impacts of loop corrections on the inflationary observables in light of the ACT results.

\subsection{Loop corrections to the inflaton potential}

We consider the effects of the loop corrections to the inflaton quartic coupling in the pole inflation scenarios.
We assume that the inflation quartic couplings in both Higgs and PQ pole inflation models run at two-loops by eq.~(\ref{twoloops}) and choose the optical renormalization scale to $\mu\sim h/\sqrt{\Omega(h)}$ or $\rho/\sqrt{\Omega(\rho)}$. 

We focus on the running Higgs quartic coupling, which becomes field-dependent, as follows,
\bea
\lambda_H(h)=\lambda_{H,e}+b_1\bigg(\ln \frac{\mu(h)}{\mu(h_e)}\bigg)+b_2\bigg(\ln \frac{\mu(h)}{\mu(h_e)}\bigg)^2
\eea
where $\lambda_{H,e}$ is the constant value of the inflaton quartic coupling at the end of inflation and
\bea
\mu(h)=\frac{h}{\sqrt{1-\frac{1}{6}h^2}}.
\eea
Then, plugging the canonical inflaton for $h$  in eq.~(\ref{can}) into the renormalization scale, we get
\bea
\ln  \frac{\mu(h)}{\mu(h_e)}&=&\ln \frac{h}{h_e} -\frac{1}{2}\ln \bigg(\frac{1-\frac{1}{6}h^2}{1-\frac{1}{6}h^2_e}\bigg) \nonumber \\
&\simeq &-2(e^{-2\phi/\sqrt{6}}-e^{-2\phi_e/\sqrt{6}})+\frac{1}{\sqrt{6}}(\phi-\phi_e) \nonumber \\
&\simeq & \frac{1}{\sqrt{6}}(\phi-\phi_e).
\eea
Here, we took $\phi,\phi_e\gtrsim \sqrt{6}$.
Therefore, instead of the tree-level inflaton potential in eq.~(\ref{tree}), we can take the loop-corrected inflaton potential, as follows,
\bea
V_{E,{\rm loops}}\simeq 9\bigg(\lambda_{H,e}+ \frac{b_1}{\sqrt{6}}(\phi-\phi_e)+ \frac{b_2}{6}(\phi-\phi_e)^2\bigg)\bigg[  \tanh\Big(\frac{\phi}{\sqrt{6}}\Big)\bigg]^{4}.
\eea
Thus, the running quartic coupling gives rise to extra contributions to the slow-roll parameters, so there is a chance to increase the spectral index towards a larger value, being consistent with ACT. For the running PQ quartic coupling, we have to replace $\lambda_{H,e}$ by $\lambda_{\Phi,e}$ and $b_{1,2}$ by the corresponding beta functions in the PQ pole inflation.

\subsection{Inflationary observables}

In the presence of the loop corrections, the slow-roll parameters get modified to
\bea
\epsilon &=&\frac{1}{2} \bigg(\frac{V'_E}{V_E}\bigg)^2 \nonumber \\
&=& \frac{1}{18} \bigg(\frac{4\sqrt{6}}{\sinh\big(\frac{2\phi}{\sqrt{6}}\big)}+\frac{3\sqrt{6}b_1+6b_2(\phi-\phi_e)}{6\lambda_{H,e}+\sqrt{6}b_1(\phi-\phi_e)+b_2(\phi-\phi_e)^2} \bigg)^2, \\
\eta &=& \frac{V^{\prime\prime}_E}{V_E} \nonumber \\
&=&\frac{2}{3}\bigg[\frac{4\Big(4-\cosh\big(\frac{2\phi}{\sqrt{6}}\big) \Big)}{\sinh^2\big(\frac{2\phi}{\sqrt{6}}\big)} +\frac{3b_2}{6\lambda_{H,e}+\sqrt{6}b_1(\phi-\phi_e)+b_2(\phi-\phi_e)^2}\nonumber \\
&&\quad+\frac{24b_1+8\sqrt{6}b_2(\phi-\phi_e)}{(6\lambda_{H,e}+\sqrt{6}b_1(\phi-\phi_e)+b_2(\phi-\phi_e)^2)\sinh\big(\frac{2\phi}{\sqrt{6}}\big)}\bigg].
\eea
We note that for $b_1=b_2=0$, the slow-roll condition is violated for $\epsilon=1$ at $\phi=\phi_e$, leading to $\sinh(\frac{2\phi_e}{\sqrt{6}})\simeq \frac{4}{\sqrt{3}}$, namely. $\phi_e\simeq 1.9$. 
Moreover, the number of efoldings is given by
\bea
N&=&\int^{\phi_*}_{\phi_e} {\rm sgn}(V'_E) \frac{d\phi}{\sqrt{2\epsilon}} \nonumber \\
&=&3\int^{\phi_*}_{\phi_e} \bigg(\frac{4\sqrt{6}}{\sinh\big(\frac{2\phi}{\sqrt{6}}\big)}+\frac{3\sqrt{6}b_1+6b_2(\phi-\phi_e)}{6\lambda_{H,e}+\sqrt{6}b_1(\phi-\phi_e)+b_2(\phi-\phi_e)^2} \bigg)^{-1}d\phi.
\eea

Assuming  the perturbative running Higgs quartic coupling, we get the slow-roll parameters evaluated at horizon exit simplified to
\bea
\epsilon_* 
&\simeq &\frac{16}{3}\sinh^{-2}\Big(\frac{2\phi_*}{\sqrt{6}}\Big) \bigg[1+\frac{3b_1+\sqrt{6}b_2 (\phi_*-\phi_e)}{12\lambda_{H,*}}\, \sinh\Big(\frac{2\phi_*}{\sqrt{6}}\Big)\bigg], \label{ep} \\
\eta_* 
&\simeq &   \frac{2}{3}\left[\frac{4\Big(4-\cosh\big(\frac{2\phi_*}{\sqrt{6}}\big) \Big)}{\sinh^2\big(\frac{2\phi_*}{\sqrt{6}}\big)} +\frac{b_2}{2\lambda_{H,*}}+\frac{4(3b_1+\sqrt{6}b_2(\phi_*-\phi_e))}{3\lambda_{H,*} \sinh\big(\frac{2\phi_*}{\sqrt{6}}\big)}\right],
\label{eta}
\eea
with $\lambda_{H,*}=\lambda_{H_e}+\frac{b_1}{\sqrt{6}}(\phi_*-\phi_e)+\frac{b_2}{6}(\phi_*-\phi_e)^2$.
Moreover, the number of efoldings also gets approximated to
\bea
N &\simeq& \frac{3}{4\sqrt{6}}\int^{\phi_*}_{\phi_e}d\phi\,\sinh\Big(\frac{2\phi}{\sqrt{6}}\Big)\bigg(1-\frac{3b_1+\sqrt{6}b_2(\phi-\phi_e)}{24\lambda_{H,*}} \, \sinh\Big(\frac{2\phi}{\sqrt{6}}\Big)\bigg) \nonumber \\
&\simeq&\frac{3}{8} \cosh\Big(\frac{2\phi_*}{\sqrt{6}}\Big)\bigg[1+\frac{b_2}{32\lambda_{H,*}}(\phi_*-\phi_e) \cosh\Big(\frac{2\phi_*}{\sqrt{6}}\Big) \nonumber \\
&&\quad-\frac{3b_1+\sqrt{6}b_2(\phi_*-\phi_e)}{48\lambda_{H,*}}  \sinh\Big(\frac{2\phi_*}{\sqrt{6}}\Big)\bigg]
\eea
or
\bea
\cosh\Big(\frac{2\phi_*}{\sqrt{6}}\Big)\simeq \frac{8}{3}N\,\bigg(1+\frac{6b_1-3b_2+2\sqrt{6}b_2 (\phi_*-\phi_e)}{36\lambda_{H,*}}\,N\bigg).
\eea

\begin{figure}[!t]
\begin{center}
 \includegraphics[width=0.45\textwidth,clip]{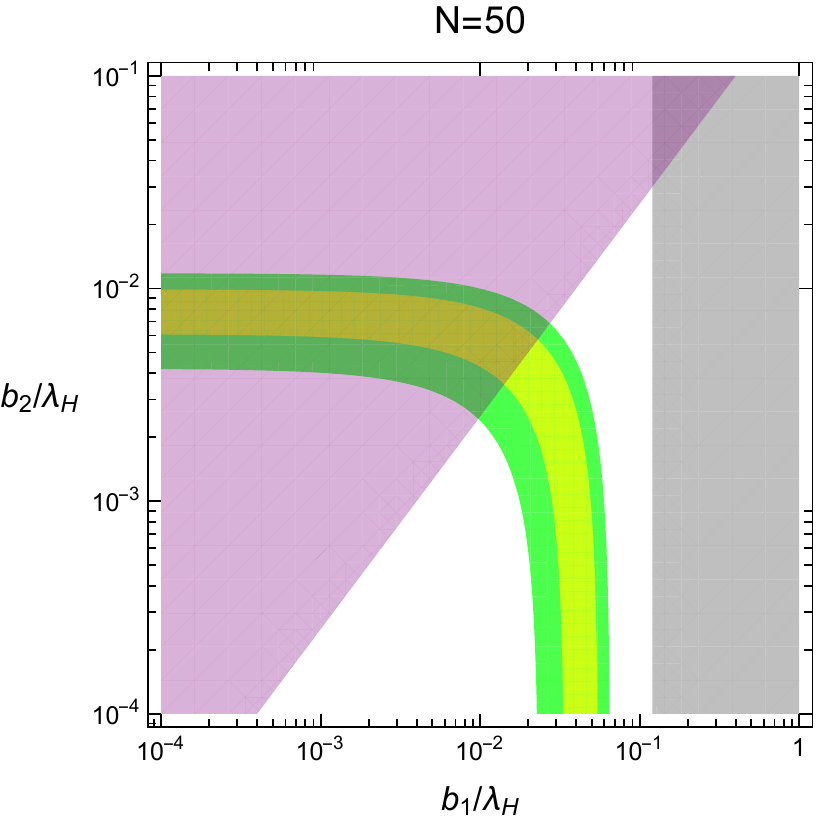}\,\, \,\, \includegraphics[width=0.45\textwidth,clip]{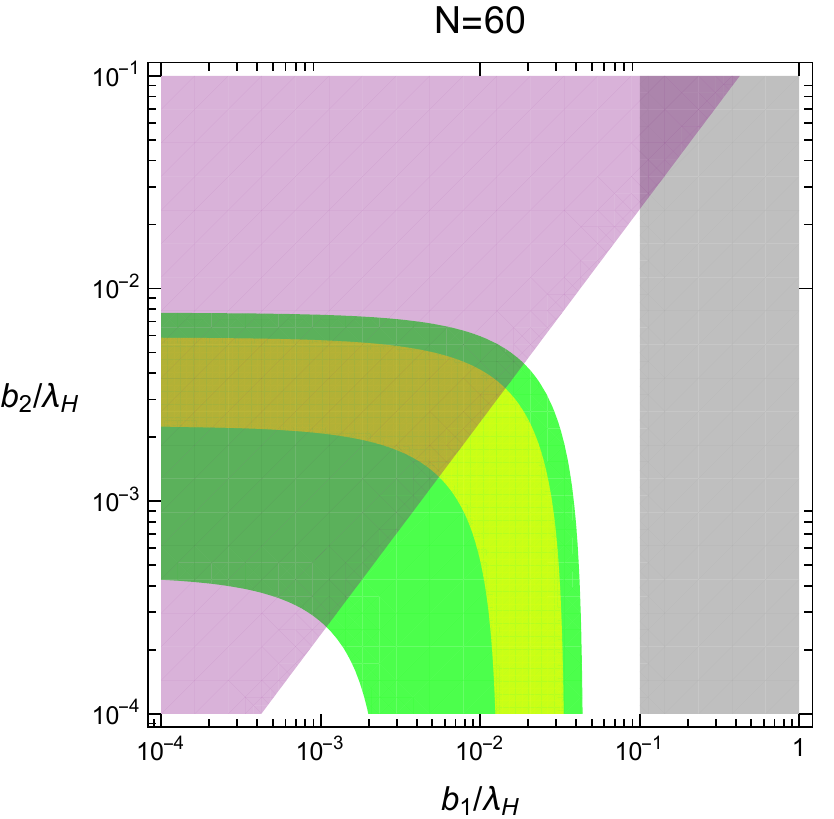}  
 \end{center}
\caption{Parameter space for $b_1/\lambda_{H_*}$ vs $b_2/\lambda_{H_*}$ being consistent with  Planck$+$ACT$+$LB data \cite{ACT}.  The range of the spectral index within $1\sigma$ and $2\sigma$ errors are shown in yellow and green. We took $N=50$ on left and $N=60$ on right. The gray region is not consistent with perturbativity and the purple region shows the two-loop dominance.}
\label{fig:para1}
\end{figure}

\begin{figure}[!t]
\begin{center}
 \includegraphics[width=0.45\textwidth,clip]{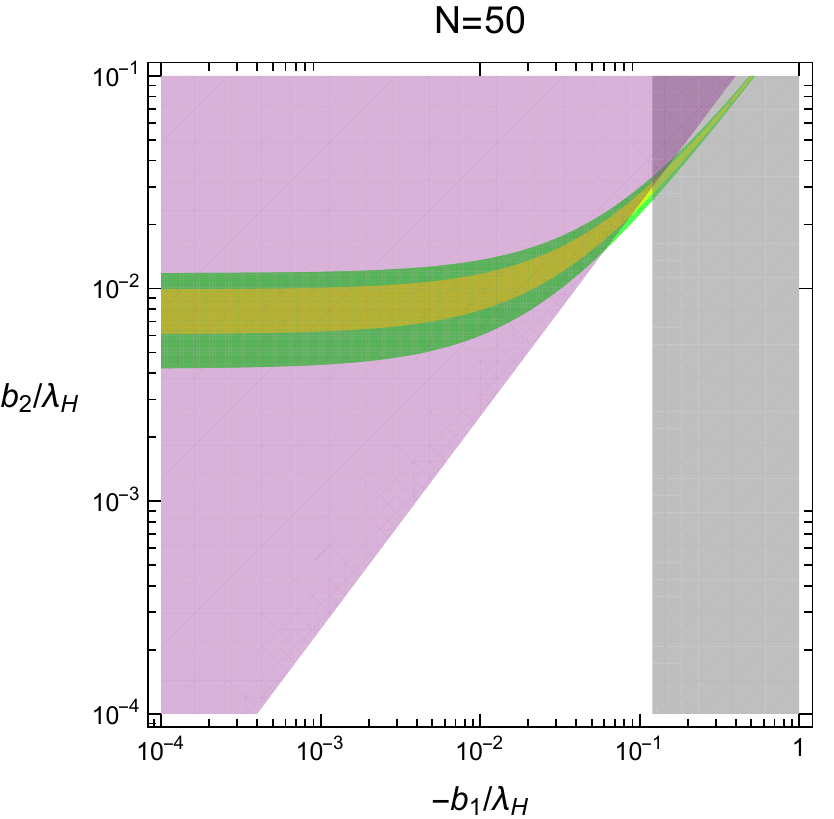}\,\, \,\, \includegraphics[width=0.45\textwidth,clip]{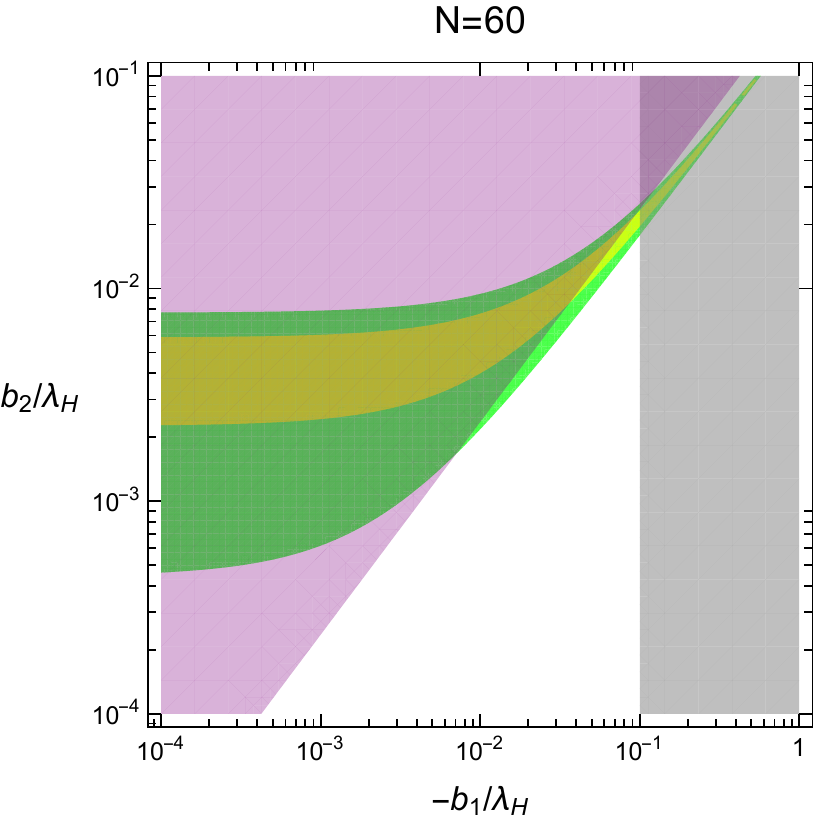}  
 \end{center}
\caption{The same as in Fig.~\ref{fig:ns}, except that $b_1$ takes negative values. }
\label{fig:para2}
\end{figure}

As a result, rewriting eqs.~(\ref{ep}) and (\ref{eta}) in terms of the number of efoldings,
\bea
\epsilon_*&\simeq& \frac{3}{4N^2}+\frac{6b_1+3b_2+2\sqrt{6} b_2 (\phi_*-\phi_e)}{24\lambda_{H,*}N}, \\
\eta_* &\simeq&-\frac{1}{N}+\frac{6b_1+9b_2+2\sqrt{6}  b_2 (\phi_*-\phi_e)}{36\lambda_{H,*}}+ \frac{6b_1+3b_2+2\sqrt{6}  b_2 (\phi_*-\phi_e) }{12\lambda_{H,*}N},
\eea
we obtain the spectral index and the tensor-to-scalar ratio at horizon exit, as follows,
\bea
n_s&=&1-6\epsilon_*+2\eta_* \nonumber \\
&\simeq&1-\frac{2}{N}+\frac{6b_1+9b_2+2\sqrt{6} b_2 (\phi_*-\phi_e)}{18\lambda_{H,*}}-\frac{6b_1+3b_2+2\sqrt{6}b_2 (\phi_*-\phi_e) }{12\lambda_{H,*}N},
\eea
and
\bea
r=16\epsilon_*\simeq\frac{12}{N^2}+\frac{2(6b_1+3b_2+2\sqrt{6} b_2 (\phi_*-\phi_e))}{3\lambda_{H,*}N}.
\eea
Therefore, $n_s$ gets corrected due to loop corrections. Under the perturbativity conditions, either signs of $b_1$ can be taken for the blue-shifted spectrum in ACT as compared to the case at tree level, but a positive $b_1$ is favored for the one-loop dominance.

The observed spectral index is given by $n_s=0.9649\pm 0.0044 $ from Planck \cite{Planck}, $n_s=0.9709\pm0.0038$ from Planck$+$ACT \cite{ACT}, and $n_s=0.9743\pm0.0034$  from Planck$+$ACT$+$LB \cite{ACT}. On the other hand, the bound on the tensor-to-scalar ratio from the combined Planck and Keck data is given by $r<0.036$ at $95\%$ CL \cite{keck}.

From the CMB normalization, $A_s=\frac{1}{24\pi^2} \frac{V_I}{\epsilon_*}=2.1\times 10^{-9}$ \cite{Planck}, with $V_I=9\lambda_{H,*}$ and $\epsilon_*=r/16$, gives rise to
\bea
\lambda_{H,*}=(3.4\times 10^{-9})\,r. \label{CMBnorm}
\eea
Thus, for a given $r$ in our model, we can determine the running quartic coupling at the horizon exit.

\begin{figure}[!t]
\begin{center}
 \includegraphics[width=0.45\textwidth,clip]{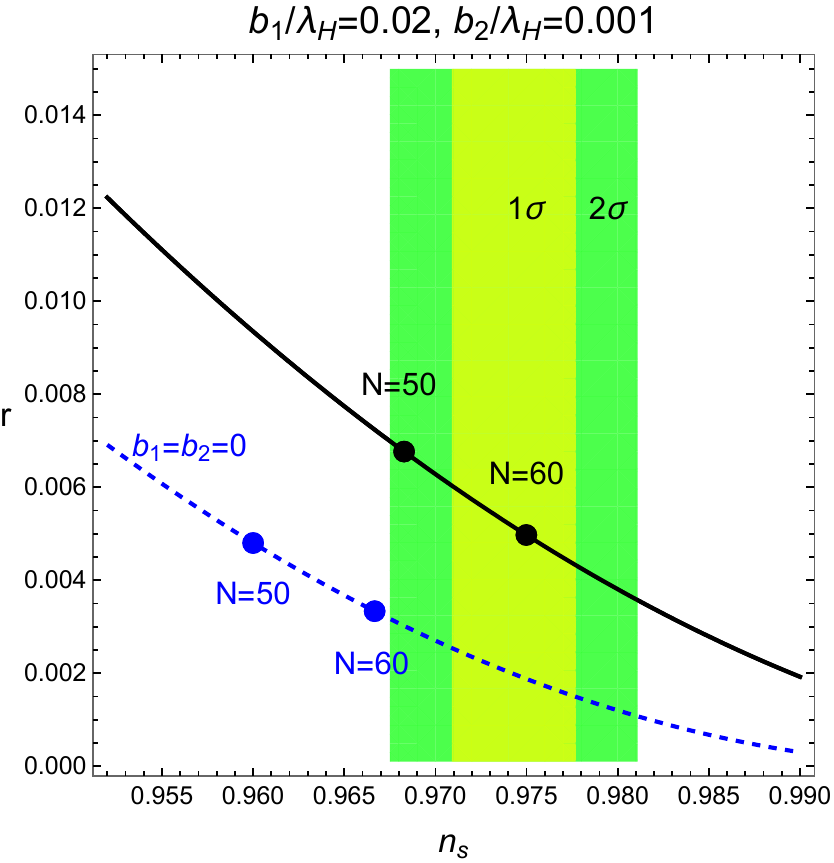}\,\, \,\, \includegraphics[width=0.45\textwidth,clip]{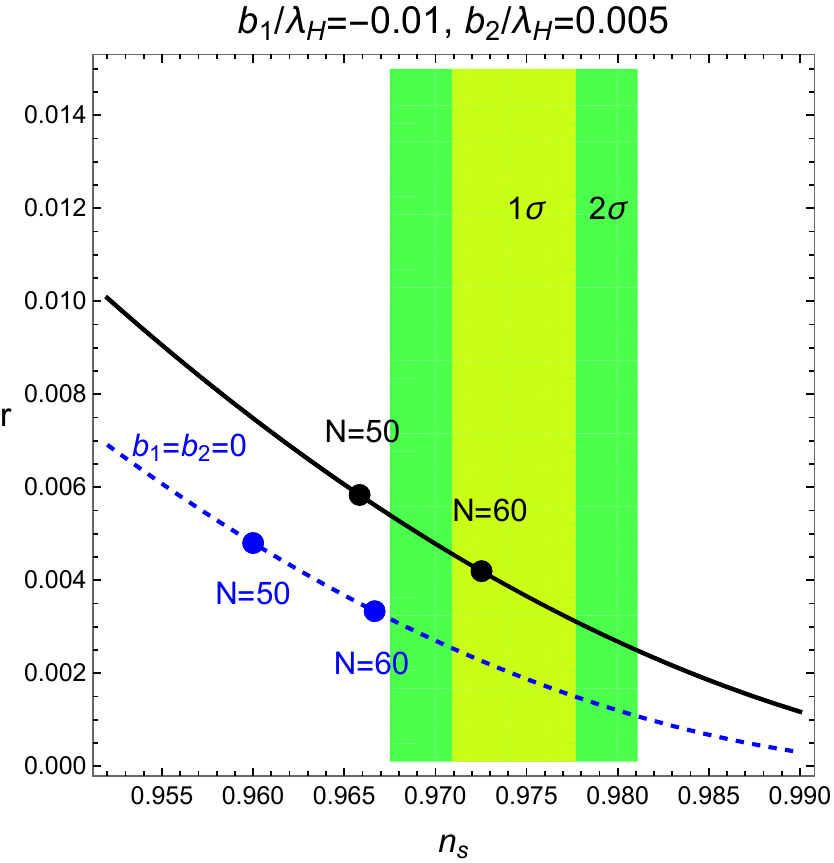}  
 \end{center}
\caption{The spectral index vs the tensor-to-scalar ratio along the sold lines in our model.  The range of the spectral index allowed by Planck$+$ACT$+$LB data \cite{ACT} are shown in yellow and green colors within $1\sigma$ and $2\sigma$ errors, respectively. We chose $b_1/\lambda_{H_*}=0.02(-0.01)$ and $b_2/\lambda_{H_*}=0.001(0.005)$ on left and right, respectively.
For comparison, in the same plots, we overlay the results at tree level for which $b_1=b_2=0$ along the dashed lines. }
\label{fig:ns}
\end{figure}

In Fig.~\ref{fig:para1}, we depict the parameter space for the one-loop and two-loop beta function coefficients, namely, $b_1/\lambda_{H_*}$ vs $b_2/\lambda_{H_*}$, that are both positive and are consistent with Planck$+$ACT$+$LB within  $1\sigma$ and $2\sigma$ errors \cite{ACT}. We took $N=50, 60$ in the left and right plots, respectively. The gray region is not consistent with perturbativity conditions for loop corrections while the purple region is where the two-loop corrections are larger than the one-loop corrections. 

In Fig.~\ref{fig:para2}, the one-loop beta function coefficient $b_1$ takes negative values. In this case, the two-loop corrections are dominant over the one-loop corrections in most of the parameter space. 

To upshot, for $b_1>0$ (singlet-dominated running) in Fig.~\ref{fig:para1},  the one-loop beta function coefficient for the Higgs quartic coupling is sufficient for explaining a larger $n_s$ as indicated by the results of ACT, because there is a reduction in the $\eta$ parameter due to one-loops only. On the other hand, for $b_1<0$ (SM-like running) in Fig.~\ref{fig:para2}, we found that the two-loop beta function coefficient needs to be as important as the one-loop beta function coefficient.

In Fig.~\ref{fig:ns}, we also show the predictions for the spectral index vs the tensor-to-scalar ratio  along the solid lines in our model. We chose the benchmark points with $b_1/\lambda_{H_*}=0.02$ and $b_2/\lambda_{H_*}=0.001$ in the left plot while $b_1/\lambda_{H_*}=-0.01$ and $b_2/\lambda_{H_*}=0.005$ in the right plot.  For $N=50-60$, the inflationary predictions can be read along the black line bounded by two black bullets.  
In these examples, for $N=50-60$, the spectral index and the tensor-to-scalar ratio vary by $n_s=0.967-0.974$ and $r=0.0050-0.0068$ on left and $n_s=0.966-0.973$ and $r=0.0040-0.0057$ on right. As a result, the running quartic coupling at the horizon exit in eq.~(\ref{CMBnorm}) varies by  $\lambda_{H,*}=(1.7-2.3)\times 10^{-11}$ on left and $\lambda_{H,*}=(1.4-2.0)\times 10^{-11}$ on right.

The results with loop corrections are in contrast with the case at tree level for $b_1=b_2=0$, as shown in dashed lines in  Fig.~\ref{fig:ns}: $n_s=0.960-0.967$ and $r=0.0033-0.0048$ for $N=50-60$, so the predictions for the spectral index are in a tension with Planck$+$ACT$+$LB at about $2-4\sigma$ level. Therefore, it is necessary to introduce the loop corrections in Higgs-like pole inflation scenarios in order to make the inflationary predictions compatible with the ACT results.

\section{Conclusions}

We presented the Coleman-Weinberg potential for the inflaton in the pole inflation scenarios through the running quartic coupling for the inflaton. 
The loop corrections stem from the SM particles and extra singlet scalar fields in the Higgs pole inflation, giving rise to the inflaton-dependent power corrections to the Higgs quartic coupling during inflation, although the non-inflaton fields are much heavier than the inflaton. Similar power corrections are obtained for the PQ pole inflation, due to the couplings of the PQ field in the realizations of the PQ symmetry: the Yukawa couplings of the heavy vector-like quark and the right-handed neutrinos and the portal coupling to the SM Higgs doublet in KSVZ models, and the Yukawa couplings of the right-handed neutrinos and the portal couplings to two Higgs doublets in DSFZ models. Thus, the concrete realizations of the pole inflation such as the beta function coefficients of the quartic coupling, $b_1$ and $b_2$, are model-dependent, but we can parametrize the running quartic coupling by the same function of the inflaton in a model-independent way, because the effective masses of the spectators fields in both models depend on the pole of the inflaton in the same way.

In the pole inflation models, there is no large parameter required for inflation, but there is a need of care in dealing with the loop corrections to the inflaton potential, coming from the SM particles as well as extra singlet scalar fields. It is necessary to introduce higher order terms in the Einstein frame potential in the renormalization process, but it was shown that the pole inflation can still work with a small quartic coupling dominating the inflaton potential. In our work, we showed that renormalization conditions are taken such that the running couplings for higher order terms in the potential, for instance, the Higgs sextet coupling, remain small and the running quartic potential becomes dominant for inflation. 

We showed that the loop corrections are necessary to make the inflationary predictions in the pole inflation deviating from the tree-level results, in favor of the ACT results. For a positive one-loop beta function for the inflaton quartic coupling (namely, $b_1>0$), a sub-dominant contribution from the two-loop corrections can be accommodated. On the other hand, if the one-loop beta function for the inflaton coupling is negative (namely, $b_1<0$), we need sizable contributions from two-loops that are larger than the one-loop corrections to explain the ACT results.
Whether $b_1>0$ or $b_1<0$, we found that the tensor-to-scalar ratio remains small enough to be compatible with the bound from Planck and Keck, although it gets slightly larger due to the loop corrections.

\section*{Acknowledgments}

The work is supported in part by Basic Science Research Program through the National
Research Foundation of Korea (NRF) funded by the Ministry of Education, Science and
Technology (NRF-2022R1A2C2003567). 
This research was supported by the Chung-Ang University Graduate Research Scholarship in 2025.

\def\theequation{A.\arabic{equation}}

\setcounter{equation}{0}

\vskip0.8cm
\noindent
{\Large \bf Appendix A: Renormalization group equations}

We present the derivation of the renormalization group (RG) equations in the Higgs pole inflation and the PQ pole inflation.

In dimensional regularization with $d=4-\epsilon$, the general form of the one-loop Coleman-Weinberg (CW) potential takes 
\bea
V_{\rm CW}= \sum_\alpha \frac{N_\alpha}{64\pi^2}\, M^4_\alpha \bigg[\ln \frac{M^2_\alpha}{\mu^2}-\frac{2}{\epsilon}+\gamma-\ln(4\pi)-C_\alpha \bigg].
\eea
In the $\overline{\rm MS}$ scheme, we need to cancel the following divergent terms by the counterterms,
\bea
V_{\rm CW, div}^{\overline{\rm MS}}=\bigg(\frac{2}{\epsilon}-\gamma+\ln(4\pi)\bigg)\sum_\alpha  \frac{N_\alpha}{64\pi^2}\, M^4_\alpha .
\eea

\noindent
\underline{Higgs pole inflation}: \\
We expand the part of the divergent terms in the CW potential in the Higgs pole inflation (with an extra singlet scalar $S$) as
\bea
\sum_\alpha  \frac{N_\alpha}{64\pi^2}\, M^4_\alpha &=& \frac{1}{64\pi^2}\bigg[ \bigg(\frac{3}{16}(g^2+g^{\prime 2})^2 +\frac{3}{8}g^4 -3y^4_t +\lambda^2_{HS}\bigg) \frac{h^4}{\big(1-\frac{1}{6}h^2\big)^2} \nonumber \\
&&+9\lambda^2_H h^4 \Big( 1-\frac{1}{6}h^2\Big)^2 \Big(1-\frac{5}{18}h^2\Big)^2 + 3\lambda^2_H h^4  \Big( 1-\frac{1}{6}h^2\Big)^2\bigg]\nonumber \\
&=&\frac{1}{64\pi^2}\bigg[ \bigg(\frac{3}{16}(g^2+g^{\prime 2})^2 +\frac{3}{8}g^4 -3y^4_t+12\lambda^2_H+\lambda^2_{HS} \bigg) h^4 \nonumber \\
&&+ \bigg(\frac{1}{16}(g^2+g^{\prime 2})^2 +\frac{1}{8}g^4 -y^4_t-9\lambda^2_H+\frac{1}{3}\lambda^2_{HS} \bigg) h^6 +\cdots\bigg] \label{expand}
\eea
where we assumed $h^2<6$ in the second line.
Thus, after cancelling the divergent terms by the counterterms,  we obtain the renormalized CW potential as 
\bea
V_R= \frac{1}{4} \lambda_H h^4 +\frac{1}{6} \lambda_6 h^6 +\dots + \sum_\alpha \frac{N_\alpha}{64\pi^2}\, M^4_\alpha \bigg[\ln \frac{M^2_\alpha}{\mu^2}-C_\alpha \bigg].
\eea
Therefore, for renormalization scale invariance, $\frac{dV_R}{d\ln\mu}=0$, the following condition must be satisfied,
\bea
\frac{1}{4}\frac{d\lambda_H}{d\ln\mu} \,h^4 +\lambda_H \frac{dh}{d\ln\mu}\,h^3 + \frac{1}{6}\frac{d\lambda_6}{d\ln\mu}\,h^6 +\lambda_6 \frac{dh}{d\ln\mu}\,h^5+\cdots-2\sum_\alpha  \frac{N_\alpha}{64\pi^2}\, M^4_\alpha=0. 
\eea
Using the above results with eq.~(\ref{expand}), we get the one-loop RG equations for the Higgs quartic coupling and the Higgs sextet coupling as
\bea
\frac{d\lambda_H}{d\ln\mu} &=& \frac{1}{8\pi^2}\bigg(\frac{3}{16}(g^2+g^{\prime 2})^2 +\frac{3}{8}g^4 -3y^4_t+12\lambda^2_H+\lambda^2_{HS} \bigg)-4\gamma_h \lambda_H, \\
\frac{d\lambda_6}{d\ln\mu} &=& \frac{3}{16\pi^2}\bigg(\frac{1}{16}(g^2+g^{\prime 2})^2 +\frac{1}{8}g^4 -y^4_t-9\lambda^2_H+\frac{1}{3}\lambda^2_{HS} \bigg)-6\gamma_h \lambda_6
\eea
where $\gamma_h$ is the one-loop wave function renormalization for the SM Higgs, given by
\bea
\gamma_h\equiv \frac{d\ln h}{d\ln\mu}=\frac{1}{64\pi^2}\Big(9g^2+3g^{\prime 2}-12y^2_t \Big).
\eea

\noindent
\underline{PQ pole inflation (KSVZ)}: \\
We expand the part of the divergent terms in the CW potential in the PQ pole inflation in KSVZ models as
\bea
\sum_\alpha  \frac{N_\alpha}{64\pi^2}\, M^4_\alpha &=&\frac{1}{64\pi^2} \bigg[ 9\lambda^3_\Phi \rho^4  \Big( 1-\frac{1}{6}\rho^2\Big)^2 \Big(1-\frac{5}{18}\rho^2\Big)^2 \nonumber \\
&&+\Big(4\lambda^2_{H\Phi} -3y^2_Q -\frac{3}{2}y^4_N\Big) \,\frac{\rho^4}{\big(1-\frac{1}{6}\rho^2\big)^2}\bigg] \nonumber \\
&=&\frac{1}{64\pi^2} \bigg[ \Big(9\lambda^2_\Phi +4\lambda^2_{H\Phi}-3y^2_Q -\frac{3}{2}y^4_N \Big)\rho^4 \nonumber \\
&&+\Big(-8\lambda^2_\Phi +\frac{4}{3}\lambda^2_{H\Phi}-y^2_Q -\frac{1}{2}y^4_N  \Big)\rho^6 +\cdots \bigg] \label{expand2}
\eea
where we assumed $\rho^2<6$ in the second line.
Thus, after cancelling the divergent terms by the counterterms,  we obtain the renormalized CW potential as 
\bea
V_R= \frac{1}{4} \lambda_\Phi \rho^4 +\frac{1}{6} {\bar\lambda}_6 \rho^6 +\dots + \sum_\alpha \frac{N_\alpha}{64\pi^2}\, M^4_\alpha \bigg[\ln \frac{M^2_\alpha}{\mu^2}-C_\alpha \bigg].
\eea
Similarly as in the Higgs pole inflation, the renormalization scale invariance leads to the following condition,
\bea
\frac{1}{4}\frac{d\lambda_\Phi}{d\ln\mu} \,\rho^4 +\lambda_\Phi \frac{d\rho}{d\ln\mu}\,\rho^3 + \frac{1}{6}\frac{d{\bar \lambda}_6}{d\ln\mu}\,\rho^6 +{\bar \lambda}_6 \frac{d\rho}{d\ln\mu}\,\rho^5+\cdots-2\sum_\alpha  \frac{N_\alpha}{64\pi^2}\, M^4_\alpha=0. 
\eea
Using the above results with eq.~(\ref{expand2}), we get the one-loop RG equations for the quartic coupling and the sextet coupling for the PQ field as
\bea
\frac{d\lambda_\Phi}{d\ln\mu} &=& \frac{1}{8\pi^2}\bigg(9\lambda^2_\Phi +4\lambda^2_{H\Phi}-3y^2_Q -\frac{3}{2}y^4_N \bigg)-4\gamma_\rho \lambda_\Phi, \\
\frac{d{\bar\lambda}_6}{d\ln\mu} &=& \frac{3}{16\pi^2}\bigg(-8\lambda^2_\Phi +\frac{4}{3}\lambda^2_{H\Phi}-y^2_Q -\frac{1}{2}y^4_N\bigg)-6\gamma_\rho {\bar\lambda}_6
\eea
where $\gamma_\rho$ is the one-loop wave function renormalization for the PQ field, given by
\bea
\gamma_\rho\equiv \frac{d\ln \rho}{d\ln\mu}=-\frac{1}{64\pi^2}\Big(12y^2_Q+6 y^2_N \Big).
\eea
We note that similar steps can be taken to derive the RG equations in DFSZ models.

\end{document}